\documentclass[runningheads]{llncs}
\usepackage[table]{xcolor}
\usepackage{stmaryrd}
\usepackage{multirow}
\usepackage{tabularx}
\usepackage{booktabs}
\usepackage{bbm}
\usepackage{breqn}
\usepackage{mathtools}
\usepackage{bm}
\usepackage{amssymb}
\usepackage{amsmath}

\usepackage{tikz}
\usepackage{subfigure}
\usetikzlibrary{backgrounds,automata}
\usetikzlibrary{bayesnet}
\usepackage{xspace}
\usepackage{amssymb}

\newcommand\iUS{\text{iUS}}

\DeclareMathOperator*{\argmin}{arg\,min}

\usepackage{xcolor,colortbl}

\definecolor{Gray}{gray}{0.85}
\definecolor{LightGray}{gray}{0.95}
\newcolumntype{a}{>{\columncolor{Gray}}c}

\usepackage{hyperref}

\begin{document}
\title{Patient-Specific Real-Time Segmentation in Trackerless Brain Ultrasound}
\titlerunning{Patient-Specific Real-Time Segmentation in Trackerless Brain Ultrasound}
\author{Reuben Dorent \inst{1} \and
Erickson Torio \inst{1} \and
Nazim Haouchine \inst{1} \and
Colin Galvin \inst{1} \and 
Sarah Frisken \inst{1} \and 
Alexandra Golby \inst{1} \and 
Tina Kapur \inst{1} \and
William Wells \inst{1,2}}
\institute{Harvard Medical School, Brigham and Women's Hospital, Boston, MA, USA  \and Massachusetts Institute of Technology, Cambridge, MA, USA
\email{rdorent@bwh.harvard.edu}
}
\authorrunning{Dorent et al}
%

%
\maketitle              
\begin{abstract}
Intraoperative ultrasound (iUS) imaging has the potential to improve surgical outcomes in brain surgery. However, its interpretation is challenging, even for expert neurosurgeons. In this work, we designed the first patient-specific framework that performs brain tumor segmentation in trackerless iUS. To disambiguate ultrasound imaging and adapt to the neurosurgeon’s surgical objective, a patient-specific real-time network is trained using synthetic ultrasound data generated by simulating virtual iUS sweep acquisitions in pre-operative MR data. Extensive experiments performed in real ultrasound data demonstrate the effectiveness of the proposed approach, allowing for adapting to the surgeon's definition of surgical targets and outperforming non-patient-specific models, neurosurgeon experts, and high-end tracking systems. Our code is available at: \url{https://github.com/ReubenDo/MHVAE-Seg}.

\keywords{Intraoperative Ultrasound \and Image Segmentation \and Cross-Modal Synthesis \and Neurosurgery}
\end{abstract}

\section{Introduction}
Intraoperative ultrasound (iUS) has recently raised considerable interest as it is an affordable, real-time, intraoperative imaging technology that can be easily integrated into existing surgical workflows.  Despite these advantages, the interpretation of iUS imaging is challenging due to the presence of artifacts, oblique orientations, reduced field of view, and low and variable contrast between different tissues, which notably complicates the identification of brain tumor margins and potential areas of residual tumor during neurosurgery~\cite{vstevno2021current}.

\begin{figure}[tp]
\begin{center}
\includegraphics[width=\textwidth]{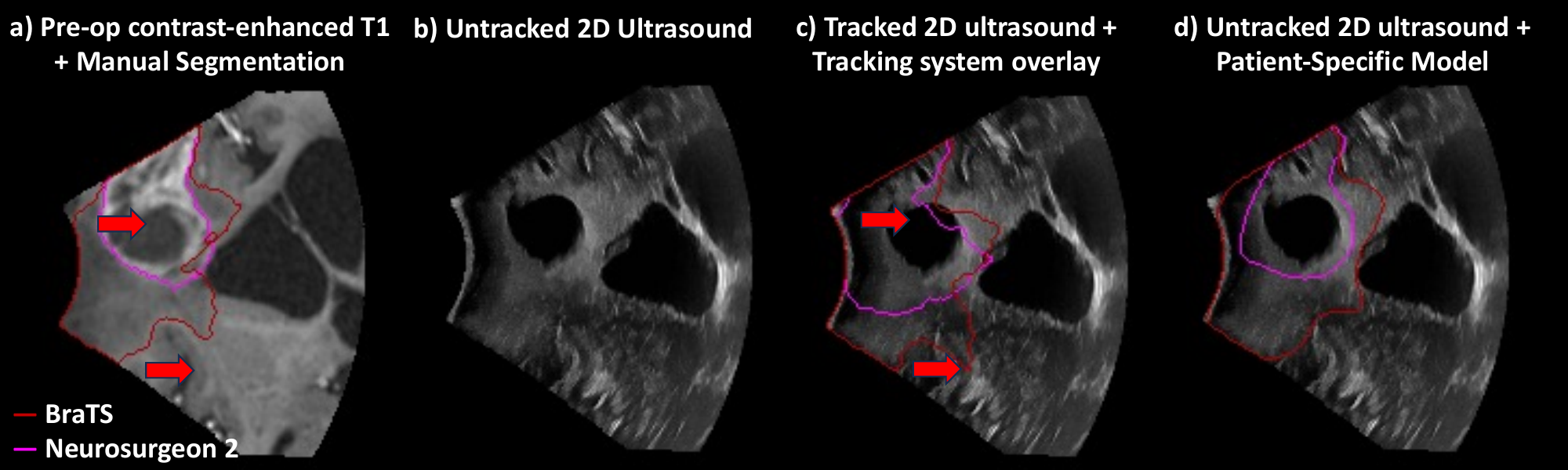}
\caption{An illustrative example showing a) the variability between neurosurgeon and BraTS annotation protocols; b) the challenge of identifying tumor boundaries in untracked 2D iUS; c) misalignment overlay issues of tracked 2D iUS with navigation system (red arrows); d) predictions of our patient-specific models in untracked 2D iUS.} \label{fig:introduction}
\end{center}
\end{figure}
 
To improve the interpretability of iUS, efforts have been made to overlay pre-operative data with iUS images using tracked iUS probes. However, relying on complex and costly navigation systems to track iUS detracts from the inherent simplicity and low cost of ultrasound imaging. Moreover, neuronavigation based on pre-operative data  becomes invalid as the surgery progresses due to brain shift and non-linear deformations caused by tissue resection, leading to misalignments between iUS and overlayed pre-operative data~\cite{cancers15030825}, as shown in Figure~\ref{fig:introduction}.
For these reasons, the use of tracked ultrasound systems remains uncommon.

Alternatively, image segmentation methods have been proposed to analyze iUS images directly~\cite{milletari2017hough,ilunga2018patient,angel2021segmentation,valanarasu2020learning,carton2020automatic}. Traditionally, this involves supervised learning with expert-annotated ultrasound images. However, segmenting ultrasound images demands rare clinical expertise and even in expert hands is subject to inter-rater variability~\cite{Weld2023.12.13.23299820}. Consequently, the size of publicly available datasets is limited~\cite{behboodi2022resect}, affecting the generalization capability and performance of these methods. Moreover, to handle data ambiguity in ultrasound imaging, segmentation often relies on 3D reconstructed images from tracking systems~\cite{ilunga2018patient,angel2021segmentation}, not native 2D iUS images, as it leads to significant performance improvements~\cite{milletari2017hough,carton2020automatic}. However, this introduces two major limitations: the reliance on navigation systems and the inability to operate in real-time due to the reconstruction process.

Another significant challenge in the surgical context is the selection of an appropriate annotation protocol. For instance, the BraTS protocol~\cite{bakas2019identifying} for brain tumors in MRI combines tumor-induced edema and infiltrated tissues into a single category, which often does not correspond to the neurosurgeon’s delineation of resectable areas. Furthermore, determining areas deemed safe for resection depends on the location of critical surrounding structures. Consequently, surgical target definitions vary across subjects and surgeons. This variability highlights the need for patient-specific segmentation strategies tailored to the patient's anatomy, and consistent with the surgeon's pre-operative planning.
 
In this work, we propose a novel segmentation approach for iUS imaging that addresses the challenges of data disambiguation and the need for patient-specificity. To our knowledge, this is the first work that segments trackerless 2D brain iUS by leveraging pre-operative data. This approach involves training a segmentation model specific to each patient by generating 2D synthetic iUS images using surgical planning data. Specifically, we assume that pre-operative MR data and the segmentation of the surgical target are available prior to surgery, which is typically the case in clinical routine.  Our contribution is three-fold. First, we designed a novel approach to generate virtual ultrasound sweeps in pre-operative MR data space, effectively mimicking the iUS acquisition process. Second, we exploit state-of-the-art Multi-Modal Hierarchical Variational Auto-Encoder~\cite{dorent2023unified} to synthesize iUS data from 2D MR slices in the virtual sweeps' field of view. To generate variability in our synthetic data, we take advantage of the stochasticity of MHVAE and its ability to handle incomplete sets of MR data. Third, extensive experiments were conducted on real ultrasound data, highlighting the need for patient-specificity and demonstrating the effectiveness of our approach by outperforming experts and tracking-system-based segmentation methods.


\section{Patient-specific segmentation in brain ultrasound}

\subsubsection{Problem overview.} The overall objective is to train a segmentation network $f_{\theta^s}$ parameterized by
weights $\theta^s$ that automatically segments 2D brain ultrasounds for a given subject~$s$. As the acquisition of adult brain ultrasound images requires skull opening (craniotomy), ultrasound images are not available prior to the surgery. Instead, we assume that we have access to a set of $M$ multi-parametric co-registered pre-operative MRI data $I^{s}=\{I_{j}^{s}\}_{j=1}^{M}$ with the segmentation of the surgical target $T^{s}$. In this work, we propose to train the patient-specific segmentation network $f_{\theta^s}$ using a set of synthetic 2D ultrasound images generated from pre-operative imaging data $(I^{s}, T^{s})$.  Our approach involves three steps: 1) simulating the positioning and trajectory of an ultrasound probe on the surface of the brain (sweep); 2) synthesizing iUS images using pre-operative MR data in the field of view of a virtual sweep; and 3) training the segmentation network $f_{\theta^s}$ using the produced synthetic iUS images and MR-based annotations.





\begin{figure}[tp]
\begin{center}
\includegraphics[width=\textwidth]{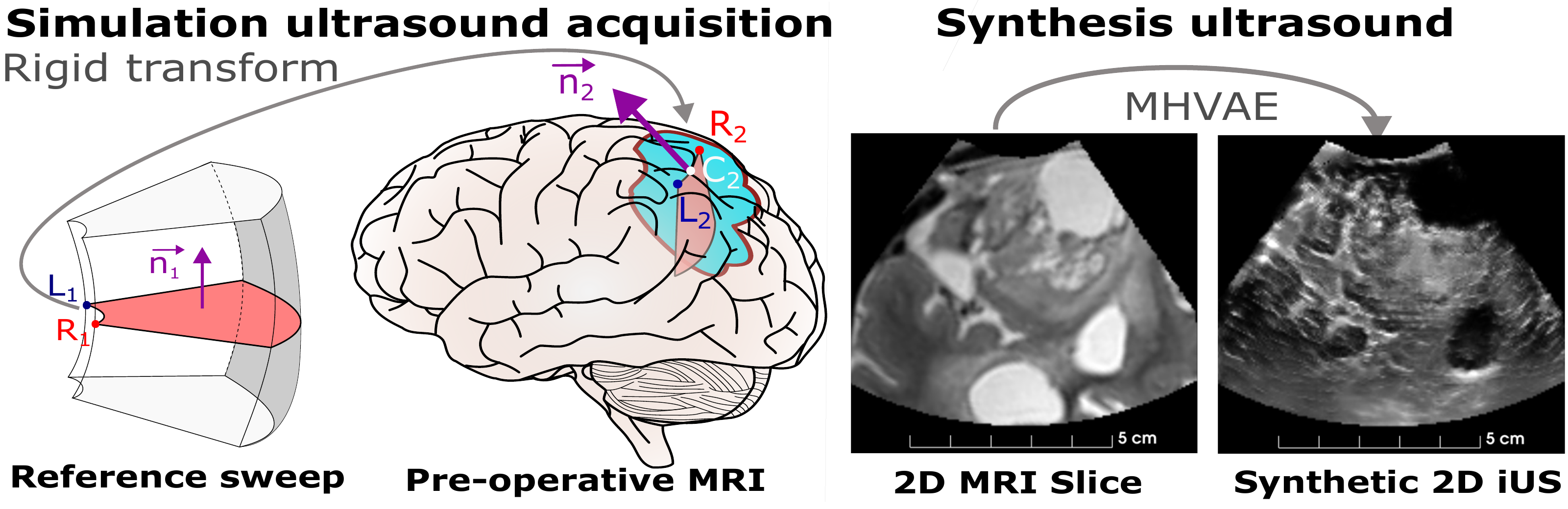}
\caption{Overview of our approach to create synthetic iUS images from pre-operative MRI. First, a generic sweep is virtually positioned on MR brain surface to simulate iUS acquisition. Then, MHVAE~\cite{dorent2023unified} synthesizes iUS from generated 2D pre-op MRI slices.}  \label{fig:simulation_synthesis}
\end{center}
\end{figure}

\subsection{Step 1: Simulation of ultrasound acquisitions}
In this section, we introduce a novel approach for simulating the acquisition of ultrasound sweeps derived from preoperative MRI data with segmented brain tumors. To accurately position the virtual iUS probe on the brain surface of our patient data $I^s$ and simulate the probe's trajectory during acquisition, we assume that: 1) the probe should be positioned on the brain surface near the tumor; 2) the sweep's trajectory should cover as much of the tumor as possible.

First, a 3D ultrasound sweep acquired during the surgery of a previous patient is randomly selected to serve as the reference for the acquisition trajectory and probe geometry. The principal trajectory direction of this reference sweep is represented by the vector $\vec{\mathbf{n_1}} \in \mathbb{R}^3$. The extremities of the probe on the median 2D slice, orthogonal to the direction of the sweep, are denoted as $\mathbf{L_1} \in \mathbb{R}^3$ and $\mathbf{R_1} \in \mathbb{R}^3$ for the left and right sides, respectively, as illustrated in Figure~\ref{fig:simulation_synthesis}. The next objective is to find a rigid transformation $R$ to place this 3D sweep in the space of the patient data $I^s$.

To ensure that the virtual sweep's trajectory covers most of the tumor, we postulate that the primary direction of the virtual ultrasound sweep in the MR space should align with the first principal component $\vec{\mathbf{n_2}} \in \mathbb{R}^3$ of the tumor's spatial distribution. Therefore, the virtual median 2D slice is within a plane $\mathcal{P}$ orthogonal to the vector $\vec{\mathbf{n_2}}$.

To fully characterize the plane $\mathcal{P}$ of the virtual median slice, an additional point $\mathbf{C}_2$ within this plane must be identified. Since the objective of the iUS acquisition is to visualize the surgical target, we propose that the plane $\mathcal{P}$ passes through a point $\mathbf{C}_2$ on the cortical surface near the surgical target. To select a candidate $\mathbf{C}_2$, we first automatically segment the brain in $I^s$~\cite{isensee2019automated} and extract the set of points on the cortical surface $\mathcal{S}=\{ \mathbf{S_i} \}$. Then, we randomly select a point \sloppy${\mathbf{C}_2 \in \mathcal{S}}$, with the probability inversely proportional to the exponential of the Euclidean distance to the tumor's centroid $\mathbf{M}_{2}$ in the segmentation $T^s$:
\begin{equation}
\mathbf{S_i} \in \mathcal{S}, \ p(\mathbf{C}_2=\mathbf{S_i}) \propto \exp(-||\mathbf{S_i}-\mathbf{M}_2||^2) \ .
\end{equation}
Thus, the virtual median 2D slice is in the plane $\mathcal{P}$ passing through $\mathbf{C}_2$ and orthogonal to $\vec{\mathbf{n_2}}$.

To accurately position the virtual median 2D slice in the plane $\mathcal{P}$, we aim to determine the optimal locations of the virtual median slice extremities $\mathbf{L_2}$ and $\mathbf{R_2}$ with three criteria: $\mathbf{L_2}$ and $\mathbf{R_2}$ should be 1/ on the cortical surface; 2/ at a similar distance to $\mathbf{C_{2}}$, 3/ their Euclidean distance should be equal to the Euclidean distance between the extremities of the median 2D slice, i.e. $\|\mathbf{L_2} - \mathbf{R_2}\|= \|\mathbf{L_1} - \mathbf{R_1}\|$. This optimization problem can be formulated as follows:
\begin{equation}
\left\{
 \begin{array}{ll}
        \mathbf{L_2} = \argmin_{\mathbf{S}_i \in \mathcal{S}} \left( \|\mathbf{S_i} - \mathbf{C_{2}} \| - 0.5 \times \|\mathbf{L_1} - \mathbf{R_1}\| \right), \\
        \mathbf{R_2} = \argmin_{\mathbf{S}_i \in \mathcal{S}} \left( \|\mathbf{S_i} - \mathbf{L_{2}} \| - \|\mathbf{L_1} - \mathbf{R_1}\| \right).
\end{array}
\right.
\end{equation}

Finally, the rigid transformation $R$ that places the reference sweep in the MR data can be estimated by aligning the sets $(\mathbf{L_1}, \mathbf{R_1}, \vec{\mathbf{n_1}})$ from the reference sweep and $(\mathbf{L_2}, \mathbf{R_2}, \vec{\mathbf{n_2}})$ from the MR data using the Least-Squares Fitting algorithm\cite{arun1987least}. 

\subsection{Step 2: Synthesizing ultrasound data from MR data}
To create synthetic MR 2D slices, we apply the rigid transformation $R$ to the reference sweep, enabling the generation of a series of 2D MR slices within the field of view of an intraoperative ultrasound probe. An illustrative example of a 2D MR slice is presented in Figure~\ref{fig:simulation_synthesis}. To synthesize iUS images from these 2D MR slices, we utilize the Multi-Modal Hierarchical Variational Auto-Encoder (MHVAE), the current state-of-the-art MRI to iUS synthesis framework~\cite{dorent2023unified}. The MHVAE architecture leverages a hierarchical latent space to capture complex and multi-level multi-modal data representations. Moreover, MHVAE has the flexibility to perform image synthesis in the presence of incomplete MR data. We use the pre-trained MHVAE model that performs iUS synthesis from any combination of the primary MR sequences for brain tumor surgery: contrast-enhanced T1 (ceT1), T2, and T2 Fluid Attenuation Inversion Recovery (FLAIR). Figure~\ref{fig:simulation_synthesis} shows an example of synthetic iUS generated using a 2D T2 slice.

To introduce variability in our synthetic data, we employ two strategies. First, we generate iUS by randomly omitting sequences from the MR data inputs. Variations in synthetic iUS are observed for different sets of input MR sequences, as shown in supplementary materials. Second, we exploit the stochastic and probabilistic nature of MHVAE to generate variability in the images. MHVAE involves sampling from Gaussian distributions at each level of the hierarchy. To introduce local variability in the generated iUS, we draw samples with various sampling temperatures. This modulates the synthetic images' local aspect, with more speckles for higher temperatures, as shown in supplementary materials.

\subsection{Step 3: Training a patient-specific segmentation network}
\subsubsection{Training set.} To train the patient-specific segmentation network, we use the two previously described steps to create a paired dataset of synthetic ultrasound with manual annotations. Given the pre-operative MR data $I^s$ and manual segmentation $T^s$, for each possible  combination of input MR images $I^s_{\pi}$ (6 combinations in total), we randomly select $K$ reference sweeps from other cases and simulate an ultrasound acquisition for each of them. Note that each acquisition simulation leads to a different positioning due to the stochastic selection of the cortical point $C_2$.  This process results in the creation of a paired set of MR 2D slices and 2D surgical target segmentations for each virtual iUS sweep acquisition. Then, we generate synthetic iUS 2D slices using MHVAE with 4 different sampling temperatures $\tau \in \{0.3, 0.5, 0.7, 1.0 \}$. In total, series of synthetic 2D iUS and surgical target annotations from $24K$ virtual sweeps are created.

\subsubsection{Network training.} Thanks to this patient-specific synthetic dataset, we can train the patient-specific segmentation network in a supervised manner. To perform real-time inference, we use a standard 2D Unet network presented in supplementary materials. Given the challenge of distinguishing tumor boundaries in iUS images, the network is deliberately trained to ``overfit" to the surgical target segmentation. However, to ensure generalization to real iUS images, extensive data augmentation techniques are implemented, including rotations, elastic deformations, scaling, mirroring, and the application of additive Gaussian noise, as well as adjustments to brightness, contrast, and gamma. At each training iteration, a complete series of synthetic iUS from a randomly selected virtual sweep is fed to the network, setting the batch size to approximately $100$ 2D images. Optimization is conducted using deep supervision with the Dice loss function, with parameters $\theta^s$ optimized through stochastic gradient descent employing Nesterov momentum (momentum set to 0.9). The model is trained for 100 epochs with an initial learning rate of 0.01, which gradually decreases to zero following a polynomial learning rate schedule. A 24GB NVIDIA A5000 GPU was used for model training.  Our code is publicly available at: \url{https://github.com/ReubenDo/MHVAE-Seg}.

\section{Experiments}
To evaluate the performance of our proposed patient-specific approach, extensive experiments were conducted on real 2D iUS images from $7$ cases  acquired during brain surgery. Three sets of brain tumor annotations  (two neurosurgeons, automated) were used for evaluation. Our patient-specific framework was compared with non-patient-specific automated methods, pre-operative segmentation overlayed by navigation systems, and manual expert iUS segmentation.

\subsubsection{Dataset.} The publicly available ReMIND dataset~\cite{remind} was used in our experiments.
$7$ cases not included during MHVAE's training were randomly selected. 
Each case comprised all pre-operative MR sequences (ceT1, T2, and FLAIR) and 3D pre-dura iUS reconstructed from sweeps of a tracked handheld 2D probe, with one case lacking the FLAIR sequence.  Note that MR images were acquired at different institutions with different protocols~\cite{remind}. The testing set corresponds to 2D slices obtained by slicing the 3D ultrasound sweeps in the direction of sweep acquisition, resulting in an average of $70$ iUS 2D slices per case.
All images were resampled to an isotropic $0.5\text{mm}$ resolution, padded for a matrix size of $(192,192)$, and normalized in $[-1,1]$. 

\begin{figure}[t]
\begin{center}
\includegraphics[width=\textwidth]{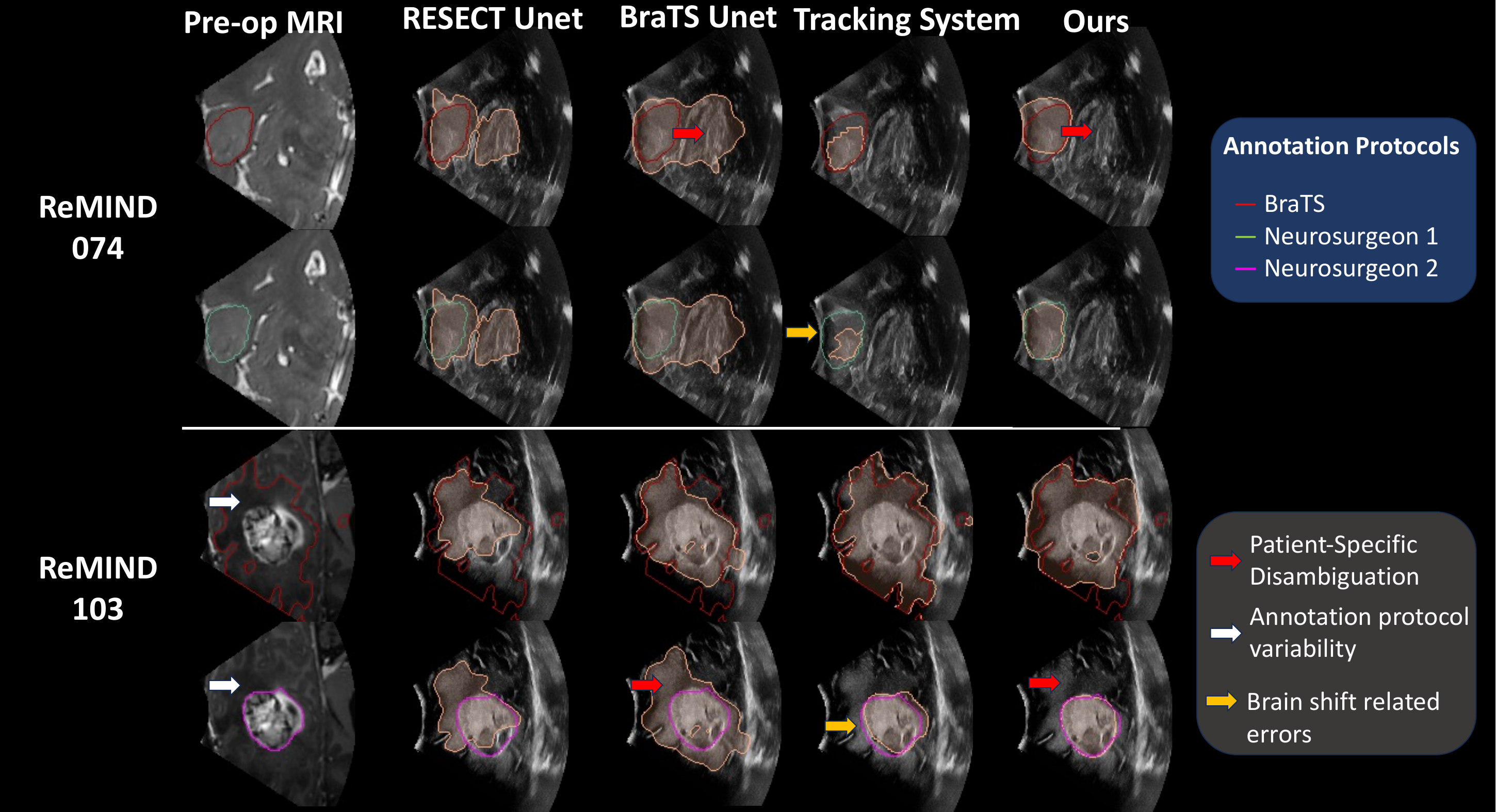}
\caption{Examples of automated surgical target segmentations using non-patient-specific models (\textit{RESECT Unet} and \textit{BraTS Unet}), high-end tracking system and \textit{Ours}. 
MR scan used for manual surgical target annotation is shown.}  \label{fig:results}
\end{center}
\end{figure} 
\subsubsection{Pre-operative segmentations.} Three sets of pre-operative MR-based segmentations were used in our experiments. Two neurosurgeons ($N_1$ and $N_2$) manually segmented the surgical target in either pre-operative ceT1 or T2 images, depending on the tumor grade and location. The segmentations by $N_1$ correspond to publicly available ReMIND annotations acquired during surgical planning.   Additionally, to simulate the \textit{BraTS} protocol, automated segmentation was conducted on ceT1 and T2 images using nnUnet~\cite{isensee2021nnu}, trained on BraTS data~\cite{bakas2019identifying}. These MR-based annotations served as the training segmentation data for our framework, which is specific to the case and the annotation protocol.

\subsubsection{Evaluation.} Since the segmentation of iUS is error-prone and ambiguous, groundtruth 2D iUS segmentations were instead obtained by propagating pre-operative MR-based segmentations via image registration between MR and 3D iUS.  This process involved affinely registering pre-operative scans with 3D pre-dura iUS using NiftyReg~\cite{MODAT2010278}, following the pipeline described in~\cite{drobny2018registration}. The registration outputs were checked by three neurological experts. To assess the accuracy of each segmentation method, Dice Similarity Coefficient (Dice) and Average Symmetric Surface Distance (ASSD)
were used, with statistical significance assessed via Wilcoxon signed-rank tests ($p < 0.01$).





\subsubsection{Determining the number of virtual sweeps.} To evaluate the impact of the total number of virtual iUS sweeps ($24K$), we evaluated the framework's performance with various $K$ values using $N_1$'s annotations. Since the duration of the framework optimization, which includes synthetic iUS dataset generation ($\approx 5\%$) and network training ($\approx 95\%$), linearly increases with $K$, the development time was tracked. As reported in Table~\ref{tab:kvalues}, the level of performance increases with the number of virtual sweeps $K$, highlighting the need for variability during training.  To balance development time with performance, we set an upper limit of 4 hours for model development, thus fixing $K=10$ in the next experiments.





\begin{table*}[t!]
    \caption{Impact of the number $K$ of generated synthetic iUS in terms of model performance and model development duration using $N_2$'s segmentations. Median and inter-quartile range are reported. Arrows
indicate favorable direction of each metric.}\label{tab:kvalues}
    \resizebox{0.99\textwidth}{!}{
	\begin{tabular}{l c *{7}{c}}
        \toprule
        & $K=1$ & $K=3$ & $K=5$ & $K=7$ & $K=10$ & $K=15$ \\
        \midrule
        \rowcolor{LightGray}
        DSC ($\%$)$\uparrow$ &  79.7 (19.4) & 78.2 (22.9) & 80.5 (19.1) & 83.6 (17.2) & 84.2 (18.1) & 84.9 (17.9) \\
        ASSD ($\text{mm}$)$\downarrow$  &2.4 (2.2) &2.4 (2.0) &2.2 (1.8) &1.9 (1.7) &1.9 (1.7) &1.8 (1.7) \\
        \rowcolor{LightGray}
        Development Duration (min)$\downarrow$ & 27 (4) & 60 (4) & 103 (8) & 148 (10) & 210 (8) & 309 (23)  \\
        \bottomrule
    \end{tabular}
    }
\end{table*}

\begin{table*}[t!]
    \caption{Quantitative evaluation of different approaches for surgical target segmentation for three annotation protocols. Median and inter-quartile range  are reported.  $^{*}$ denotes significant improvement provided by a Wilcoxon test ($p < 0.01$). }\label{tab:results}
    \resizebox{0.99\textwidth}{!}{
	\begin{tabular}{l c *{7}{c}}
        \toprule
        & \multicolumn{2}{c}{\bf BraTS  }  & \multicolumn{2}{c}{\bf Neurosurgeon 1 }  & \multicolumn{2}{c}{\bf Neurosurgeon 2 }   \\
        & \multicolumn{2}{c}{\bf (487 2D iUS slices) }  & \multicolumn{2}{c}{\bf (487 2D iUS slices)}  & \multicolumn{2}{c}{\bf (70 2D iUS slices) }   \\
        \cmidrule(lr){2-3} \cmidrule(lr){4-5} \cmidrule(lr){6-7} 
        & DSC ($\%$)$\uparrow$  & ASSD ($\text{mm}$)$\downarrow$ & DSC ($\%$)$\uparrow$  & ASSD ($\text{mm}$)$\downarrow$ & DSC ($\%$)$\uparrow$  & ASSD ($\text{mm}$)$\downarrow$ \\
        \midrule
        \rowcolor{LightGray}
        RESECT Unet &71.7 (17.4) &4.2 (3.5) &59.8 (33.2) &5.4 (3.5) &58.5 (30.4) &5.3 (2.8) \\
        Brats Unet &76.8 (12.8) &3.9 (2.4) &56.1 (28.8) &6.0 (3.1) &53.8 (30.2) &5.8 (3.8) \\
        \rowcolor{LightGray}
        Tracking System &85.4 (9.2) &\textbf{1.9* (0.7)} &80.0 (16.9) &2.2 (1.4) &80.1* (16.4) &2.2* (1.0) \\
        Manual &  $\times$ & $\times$ & $\times$ &  $\times$ & 67.4 (29.4) &3.0 (2.2) \\
        \rowcolor{LightGray}
        \midrule
        \textbf{Ours (K=10)} & \textbf{87.2* (8.5)} &2.1 (1.7) &\textbf{84.2* (18.1)} &\textbf{1.9* (1.7)} &\textbf{84.2* (36.6)} &\textbf{1.8* (2.4)} \\
        \bottomrule
    \end{tabular}
    }
\end{table*}

\subsubsection{Baseline approaches.} We compared our patient-specific approach against three baselines. First, we trained two non patient-specific 2D Unet approaches:  \textit{RESECT Unet}, using the only publicly available dataset of brain iUS with manual tumor annotations (RESECT-SEG~\cite{behboodi2022resect}, $N=23$ 3D \iUS) and \textit{BraTS Unet}, using a synthetic dataset of iUS generated by applying our proposed method to a large, annotated subset of the BraTS MRI dataset  (UPenn-GBM~\cite{bakas2022university}, $N=611$ cases). Second, we compared our trackless method to the patient-specific segmentation derived from the high-end optical tracking of the ``Curve'' navigation system (Brainlab AG). Third, an experienced neurosurgeon ($N_2$) with expertise in iUS imaging manually segmented the surgical target in preoperative MRI and subsequently in $10$ random iUS slices per case. 

\subsubsection{Results.} 
Quantitative and qualitative results are shown in Table~\ref{tab:results} and Figure~\ref{fig:results}. Firstly, we can observe that \textit{RESECT Unet} and \textit{BraTS Unet} obtained comparable performance on the ReMIND dataset, showing that our framework circumvents the need for iUS manual annotations. Secondly, the comparison between \textit{Ours} and \textit{BraTS Unet} highlights the need for patient-specificity. On the one hand, leveraging pre-operative MR data allows disambiguating iUS images, as shown in Figure~\ref{fig:results}, significantly increasing the Dice by $+11.4$pp  and reducing the ASSD by a factor of $2$ for BraTS annotations. On the other hand, the performance of \textit{BraTS Unet} drops when compared with $N_1$ and $N_2$'s annotations, illustrating the variability in annotation protocols. In contrast, our patient-specific approach obtained stable results across protocols, demonstrating the advantage of leveraging neurosurgeon pre-operative planning during training. Third, our trackerless approach surprisingly reached comparable performance to the tracking-system-based method, and even outperformed it in some cases. This suggests our approach may provide a viable alternative to expensive and complex tracking systems without compromising performance. Finally, the mediocre results achieved by the iUS expert emphasize the inherent challenges of segmenting brain surgical targets in iUS.  Notably, our approach significantly outperforms the expert neurosurgeon, suggesting its potential utility in assisting surgeons. Overall these experiments demonstrate the effectiveness of our approach, which leverages a real-time 2D Unet ($200$ FPS) and pre-operative data, to create a patient-specific segmentation model adapted to the neurosurgeon.

\section{Conclusion}
In this work, we introduced a novel approach to design patient-specific models on the challenging problem of surgical target segmentation in brain intraoperative ultrasound. The proposed framework can be seen as a patient-specific cross-modality domain adaptation approach~\cite{dorent2023crossmoda}. Our approach, which only requires access to pre-operative planning, could be a viable alternative to complex tracking systems to assist with iUS interpretation. In line with recent medical imaging research, this work advocates for patient-specific models leveraging pre-operative data to improve performance during surgery. Future work will expand automated segmentation to additional brain structures, incorporate temporal consistency during iUS sweeps acquisition and focus on reducing the development duration. From an application perspective, we will explore whether our framework can automatically detect significant misalignments between pre-operative and intra-operative data. This could alert surgeons to inaccuracies in the navigation system. 

\subsubsection{Acknowledgement}
This work was supported by the National Institutes of Health (R01EB032387, R01EB027134, P41EB015902, P41EB028741 and K25EB035166). We also thank the philanthropic support from the Jennifer Oppenheimer Cancer Research Initiative and Karen and Brian McMahon.

\bibliographystyle{splncs04}
\bibliography{paper2711.bib}

\newpage


\section{Supplementary materials}

\begin{figure}[b!]
\begin{center}
\includegraphics[width=\textwidth]{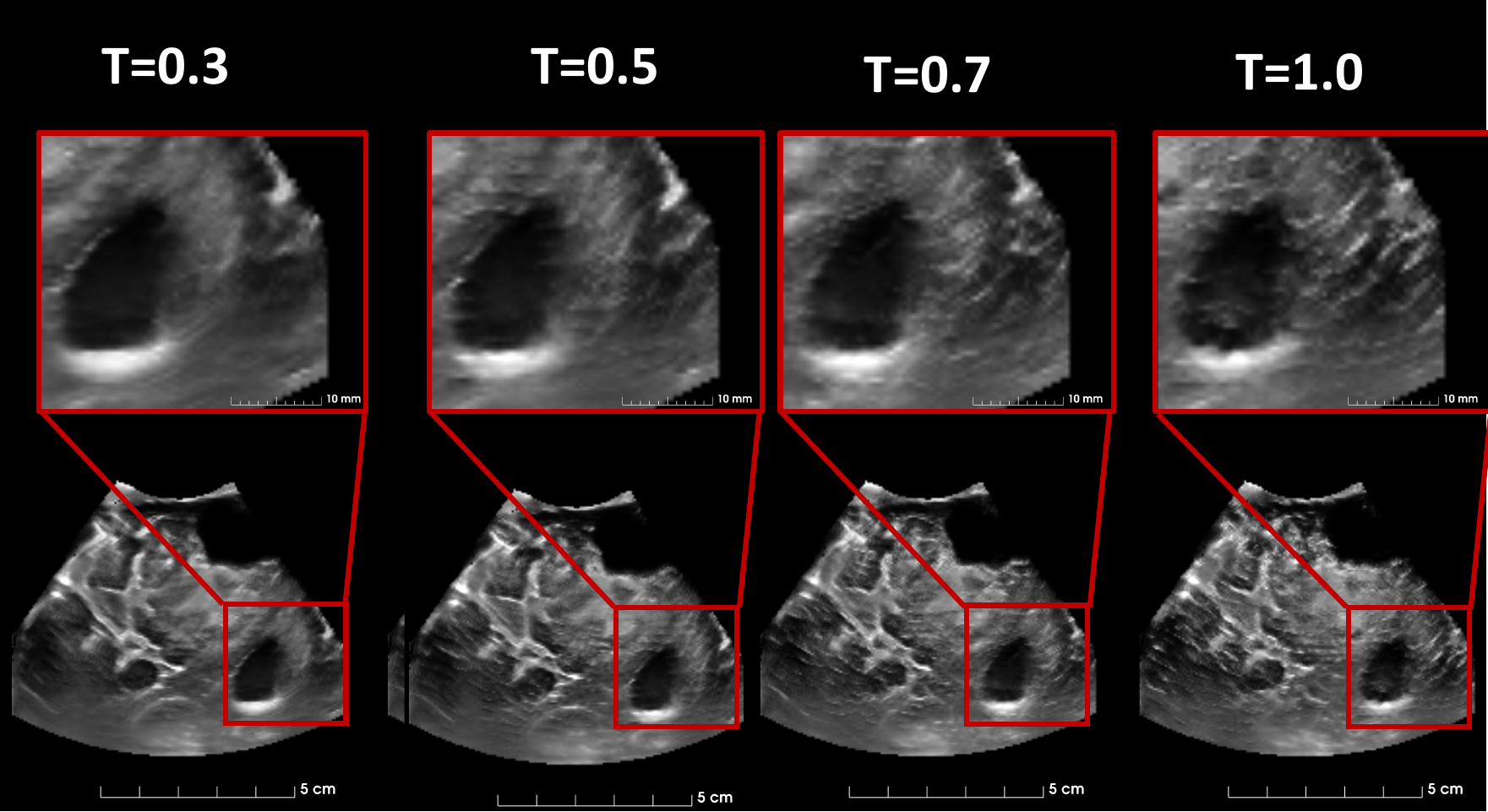}
\caption{Impact of the temperature $\tau$ on the local variability (speckles).} 
\end{center}
\end{figure}

\begin{figure}[b!]
\begin{center}
\includegraphics[width=\textwidth]{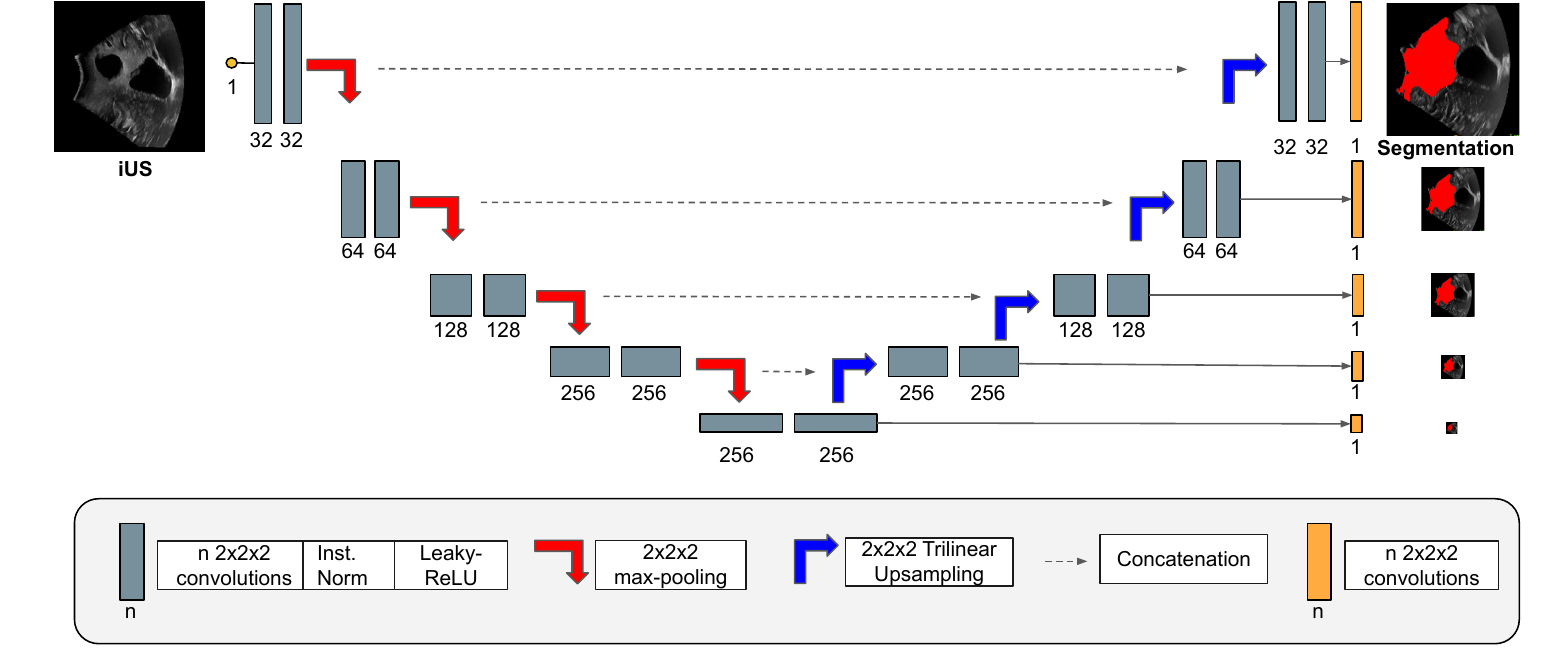}
\caption{Our 2D Unet segmentation network.} 
\end{center}
\end{figure}

\end{document}